\documentclass[aps,prl,floatfix,twocolumn,superscriptaddress]{revtex4-2}

\usepackage{graphicx,
            caption,
            subcaption,
            amsmath,
            braket,
            units,
            ragged2e,
            xcolor,
            xspace,
            nicefrac,
            lipsum,
            nameref,
            textcomp,
            urwchancal,
            upgreek,
            isotope,
            hyperref,
            cleveref,
            siunitx,
            placeins, 
            scrextend, 
            url}

\DeclareCaptionLabelSeparator{dot}{. }
\makeatletter
\def\justified{
	\let\\\@normalcr
	\@rightskip\z@skip \rightskip\@rightskip
	\leftskip\z@skip
	\parindent 0em\relax
	\setlength{\parfillskip}{0pt plus 1fil}}
\DeclareCaptionJustification{justified}{\justified}
\hypersetup{colorlinks=true, linkcolor=blue,citecolor=blue,urlcolor=blue}
\def\unit #1 #2 {\SI{#1}{#2}\xspace}
\def\range #1 #2 #3 {\SIrange{#1}{#2}{#3}\xspace}
\DeclareSIUnit\gauss{G}

\newcommand{\myref}[2][]{Fig.~\hyperref[#2]{\ref*{#2}#1}}
\newcommand{\Myref}[2][]{Figure~\hyperref[#2]{\ref*{#2}#1}}
\newcommand{\Mytabref}[2][]{Table~\hyperref[#2]{\ref*{#2}#1}}

\newcommand{\am}{A_\text M}
\newcommand{\aphi}{A_\Phi}

\newcommand{\ii}{\text i}
\newcommand{\tho}{t_h}

\newcommand{\ie}{i.\,e.}
\newcommand{\eg}{e.\,g.}

\begin{document}

\title{Birth, life, and death of a dipolar supersolid}

\date{\today}

\author{Maximilian Sohmen}
\affiliation{
    Institut f\"{u}r Quantenoptik und Quanteninformation, \"Osterreichische Akademie der Wissenschaften, Innsbruck, Austria
}
\affiliation{
    Institut f\"{u}r Experimentalphysik, Universit\"{a}t Innsbruck, Austria
}

\author{Claudia Politi}
\affiliation{
    Institut f\"{u}r Quantenoptik und Quanteninformation, \"Osterreichische Akademie der Wissenschaften, Innsbruck, Austria
}
\affiliation{
    Institut f\"{u}r Experimentalphysik, Universit\"{a}t Innsbruck, Austria
}

\author{Lauritz Klaus}
\affiliation{
    Institut f\"{u}r Quantenoptik und Quanteninformation, \"Osterreichische Akademie der Wissenschaften, Innsbruck, Austria
}
\affiliation{
    Institut f\"{u}r Experimentalphysik, Universit\"{a}t Innsbruck, Austria
}

\author{Lauriane Chomaz}
\affiliation{
    Institut f\"{u}r Experimentalphysik, Universit\"{a}t Innsbruck, Austria
}

\author{Manfred J. Mark}
\affiliation{
    Institut f\"{u}r Quantenoptik und Quanteninformation, \"Osterreichische Akademie der Wissenschaften, Innsbruck, Austria
}
\affiliation{
    Institut f\"{u}r Experimentalphysik, Universit\"{a}t Innsbruck, Austria
}

\author{Matthew A. Norcia}
\affiliation{
    Institut f\"{u}r Quantenoptik und Quanteninformation, \"Osterreichische Akademie der Wissenschaften, Innsbruck, Austria
}
	
\author{Francesca Ferlaino}
\thanks{Correspondence should be addressed to \mbox{\url{Francesca.Ferlaino@uibk.ac.at}}}
\affiliation{
    Institut f\"{u}r Quantenoptik und Quanteninformation, \"Osterreichische Akademie der Wissenschaften, Innsbruck, Austria
}
\affiliation{
    Institut f\"{u}r Experimentalphysik, Universit\"{a}t Innsbruck, Austria
}


\begin{abstract}
In the short time since the first observation of supersolid states of ultracold dipolar atoms, substantial progress has been made in understanding the zero-temperature phase diagram and low-energy excitations of these systems.
Less is known, however, about their finite-temperature properties, particularly relevant for supersolids formed by cooling through direct evaporation.  
Here, we explore this realm by characterizing the evaporative formation and subsequent decay of a dipolar supersolid by combining high-resolution in-trap imaging with time-of-flight observables.  
As our atomic system cools towards quantum degeneracy, it first undergoes a transition from thermal gas to a crystalline state with the appearance of periodic density modulation.  This is followed by a transition to a supersolid state with the emergence of long-range phase coherence.  
Further, we explore the role of temperature in the development of the modulated state.  
\end{abstract}

\maketitle

Supersolid states, which exhibit both global phase coherence and periodic spatial modulation~\cite{Gross:1957,Gross:1958,Andreev:1969,Chester:1970,Leggett:1970, Li:2017, Leonard:2017}, have recently been demonstrated and studied in ultracold gases of dipolar atoms~\cite{Bottcher2019,Tanzi:2019,Chomaz:2019}.  
These states are typically accessed by starting with an unmodulated Bose--Einstein condensate (BEC), and then quenching the strength of interatomic interactions to a value that favors a density-modulated state. 
In this production scheme, the superfluidity (or global phase coherence) of the supersolid is inherited from the pre-existing condensate. 
However, a dipolar supersolid state can also be reached by direct evaporation from a thermal gas with fixed interactions, as demonstrated in Ref.\,\cite{Chomaz:2019}.

A thermal gas at temperatures well above condensation has neither phase coherence nor modulation, so both must emerge during the evaporative formation process. 
This leads one to question whether these two features appear simultaneously, or if not, which comes first.  
Further, because this transition explicitly takes place at finite temperature $T$, thermal excitations may play an important role in the formation of the supersolid, presenting a challenging situation for theory.  
Moreover, in the case of a dipolar supersolid, the non-monotonic dispersion relation and the spontaneous formation of periodic density modulation lead to important new length- and energy-scales not present in contact-interacting systems, which dramatically modify the evaporative formation process. 

\begin{figure}[ht]
    \centering
	\includegraphics[width=\columnwidth]{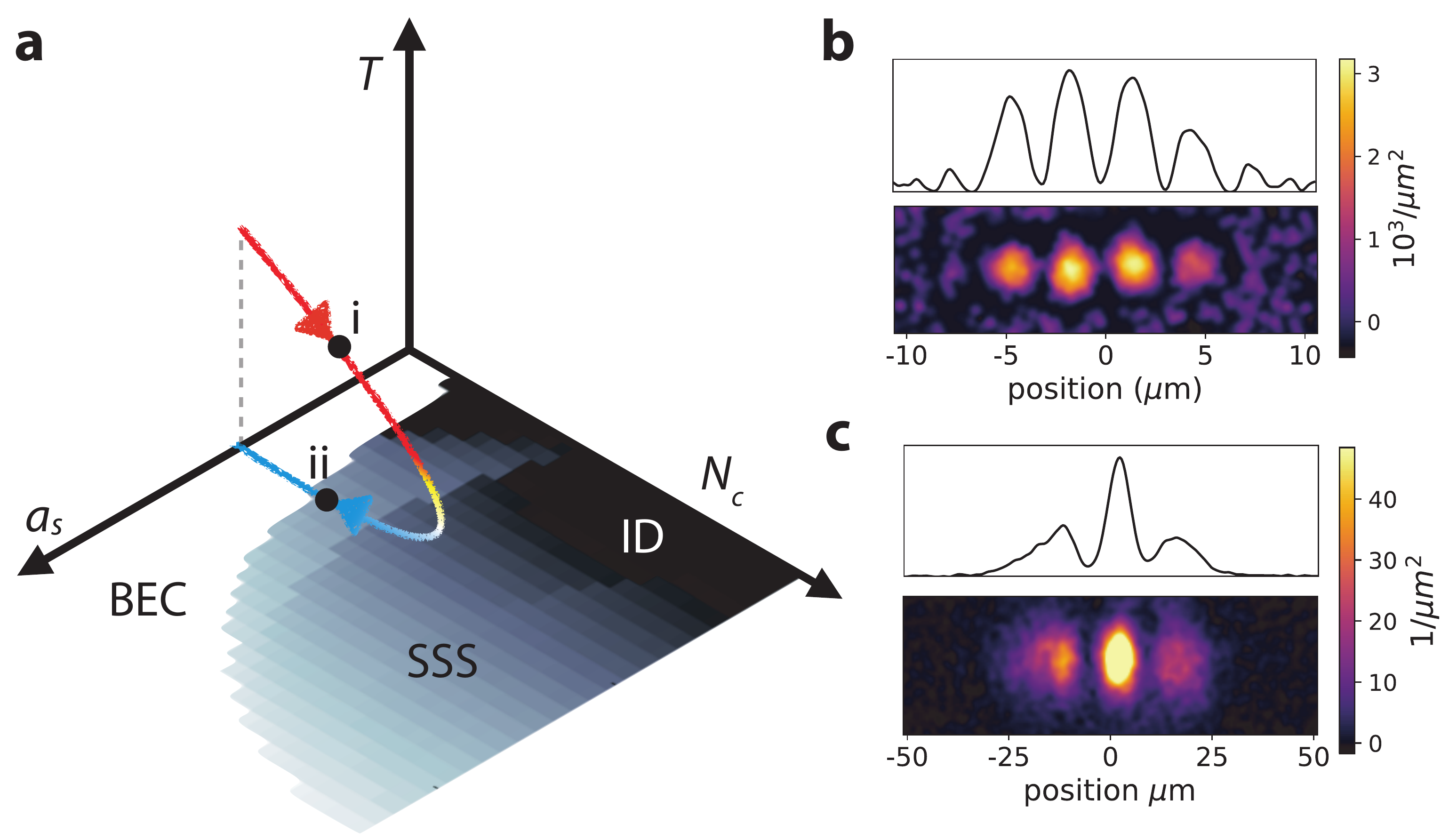}
	\caption {
	{\bf Evaporation trajectory through the finite-temperature phase diagram.}
	\textbf{a.}  
	At $T = 0$ (bottom plane), the phase diagram for a gas of dipolar atoms is spanned by the s-wave scattering length $a_s$ and the condensate atom number $N_c$.
	In an elongated trap it features a BEC (white) and independent droplet (ID, black) phases, separated in places by a supersolid state (SSS, gray-scale).  
	The plotted lightness in the $T=0$ phase diagram represents the droplet link strength across the system (cf.~\cite{Ilzhofer:2019}).  
	Away from $T=0$, the phase diagram is not known.  
	We explore this region through evaporation into (red, near i) and out of (blue, near ii) the SSS, along a trajectory represented schematically by the colored arrow.  
	\textbf{b.}	Single-shot image of the optical density (OD) of the sample in trap.  
	Here, a system of four ``droplets" within the SSS region is shown, together with its projected density profile.  
	\textbf{c.} Single-shot matter-wave interference pattern after 35\,ms TOF expansion (OD), and the corresponding projected profile. 
	The color-scale is truncated for visual clarity.  
    The background clouds of thermal atoms present are not visible in the color scales of subfigures b,\,c;  
    for $35$\,ms TOF and around $\SI{50}{\nano\kelvin}$ (as in c) the thermal atoms show an approximately isotropic 2D Gaussian distribution of width $\bar\sigma\sim\SI{55}{\micro\meter}$.}
	 \label{fig:1} 
\end{figure}

While the ground state and dynamics of a zero-temperature dipolar quantum gas can be computed by solving an extended Gross--Pitaevskii equation~\cite{Waechtler:2016,Chomaz:2016,Bisset:2016,Waechtler:2016b,Baillie:2018,Bottcher2019,Ilzhofer:2019,Roccuzzo:2019} (see also Fig.\,\ref{fig:1}a), similar treatments are currently lacking for finite temperatures in the supersolid regime.  
In principle, effects of finite temperature can be taken into account by perturbatively including the thermal population of excited modes.  This can be done either coherently, by adding them in a single classical field which abides the Gross--Pitaevskii equation, as in Refs.~\cite{blakie2008dynamics,Petter:2020, Hertkorn:2020}, or incoherently, by iteratively computing mode populations via a set of coupled Hartree--Fock--Bogoliubov equations~\cite{Ronen2007,Aybar2019,Tanzi:2019}.  In order to accurately describe dynamical processes occurring at temperatures approaching the critical temperature, both coherent excitations and incoherent interactions with the background thermal gas must be accounted for, requiring either more advanced c-field~\cite{blakie2008dynamics} or quantum Monte Carlo~\cite{Cinti2010sdc,Saito2016,Macia:2016,Cinti:2017,kora2019patterned} techniques.  
So far, theories with realistic experimental parameters have not been developed to unveil the finite-temperature dipolar phase diagram and to determine the properties of the thermal-to-supersolid phase transitions.

In this work, we experimentally study the evaporative transition into and out of a supersolid state in a dilute gas of dysprosium atoms.  
As the atoms cool down to quantum degeneracy, the number of condensed atoms increases, giving birth to the supersolid state. Continued evaporation and collisional loss lead to a reduction of atom number, and eventually the death of the supersolid.
Such an evaporation trajectory, as illustrated in Fig.~\ref{fig:1}a, passes through the little-understood finite-temperature portion of the supersolid phase diagram.  
During the evaporative birth of the supersolid, we discover that the system first establishes strong periodic density modulation of locally coherent atoms, and only later acquires long-range phase coherence.  
When comparing the birth and death of the supersolid, which occur at different temperatures, we observe higher levels of modulation during the birth, suggesting that thermal fluctuations may play an important role in the formation of density modulation.

For our experiments, we first prepare an optically trapped gas of approximately $10^5$ dysprosium atoms (isotope $^{164}$Dy), precooled via forced evaporation to temperatures of several hundred nanokelvin, at which point the gas remains thermal.  
From here, we can apply further evaporation either by a nearly-adiabatic  ramp-down of the trap depth (``slow ramp"), or by a rapid reduction of the trap depth followed by a hold time at fixed depth (``fast ramp'') to further lower the temperature and induce condensation into the supersolid state. The ``slow ramp'' protocol yields a higher number of condensed atoms ($N_c \sim 2 {\times} 10^4$; see next paragraph for definition), and lower shot-to-shot atom number fluctuations, whereas the ``fast ramp'' protocol ($N_c \sim  10^4$) allows to follow the evolution of the system in a constant trap, disentangling the system dynamics from varying trap parameters.  
In contrast to protocols based on quenching the interactions in a BEC~\cite{Tanzi:2019,Bottcher2019,Chomaz:2019}, we hold the magnetic field (and hence the contact interaction strength) fixed during the entire evaporation process at 17.92\,G, where the system ground state at our $N_c$ is a supersolid (scattering length $\sim 85(5)$\,$a_0$).
\\

For the present work, we have implemented in-situ Faraday phase contrast imaging~\cite{Bradley:1997,Kadau:2016}, which allows us to probe the in-trap density of our quantum gas at micron-scale resolution. 
During the formation of the density-modulated state, the translation symmetry is broken along the long (axial) direction of our cigar-shaped trap~%
\footnote{\label{note:adiabatic} At the end of evaporation, we typically use trap frequencies around $\omega_{x,y,z} = 2\pi{\times}(36,88,141)\,\text s^{-1}$, with the tightest trap direction oriented parallel to gravity and to an applied uniform magnetic field, cf.~\cite{Chomaz:2019}.}%
, typically giving rise to a chain of three to six density peaks, which we call ``droplets."
These droplets have a spacing of roughly three microns, clearly visible in our in-situ images (Fig.~\ref{fig:1}b).   
As in our previous works~\cite{Chomaz:2019,Ilzhofer:2019}, we also image the sample after a time-of-flight (TOF) expansion using standard absorption imaging.  
These TOF images include a spatially broad contribution which we attribute to thermal atoms, whose number $N_{th}$ and temperature $T$ we estimate by 2D-fitting of a  Bose-enhanced Gaussian function~\cite{Ketterle:1999}, excluding the cloud centre.  
Surplus atoms at the cloud centre (compared to the broad Gaussian) are at least locally coherent, or ``\mbox{(quasi-)condensed}'' in the sense of Refs.~\cite{Petrov:2001,Dettmer:2001,prokof2004weakly}.  
With the total number of atoms $N$ measured by pixel count, we define $N_c = N - N_{th}$ to be the number of these (at least locally) coherent atoms.  
During TOF, matter-wave interference between the expanding droplets gives rise to a characteristic interference pattern (Fig.~\ref{fig:1}c). 
The high contrast of the interference pattern is visible in single TOF images and indicates that each individual droplet is by itself a phase coherent many-body object.
The stability of the interference fringes within the envelope over multiple experimental realisations encodes the degree of phase coherence between droplets~(cf.~Refs.~\cite{Chomaz:2019,Ilzhofer:2019} and discussion below). 
The combination of in-situ and TOF diagnostics provides complementary information allowing us to measure both density modulation and its spatial extent (number of droplets), as well as phase coherence.  

\begin{figure}[ht]
    \centering
	\includegraphics[width=\columnwidth]{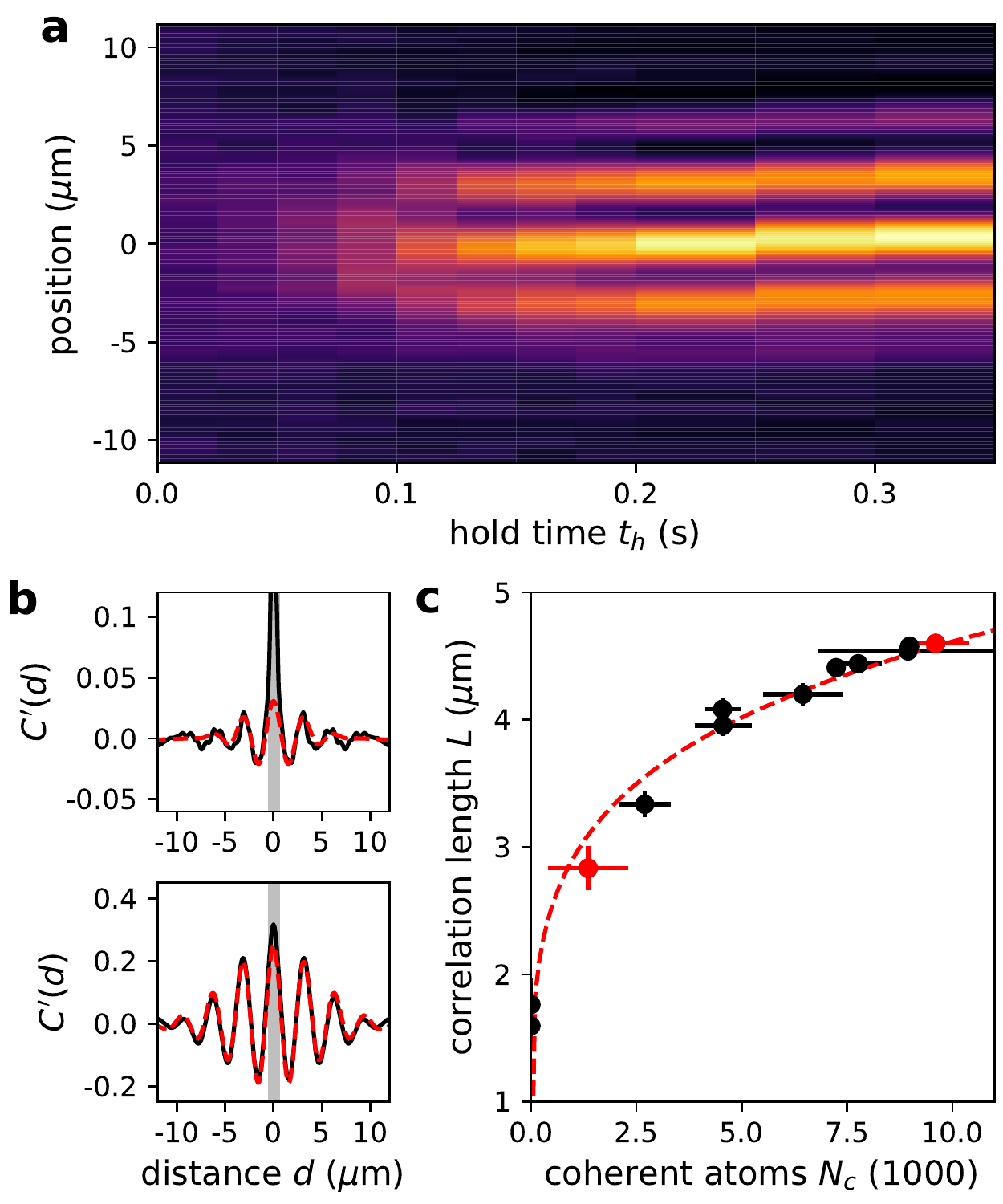}
	\caption {
	{\bf Growth and spread of density modulation during evaporation.}
	{\bf a.} Averaged density profiles (no recentering, approximately 20 shots per timestep) along the long trap axis as a function of hold time $t_h$ after the ``fast ramp'' reduction of trap depth (see main text).
	{\bf b.} The density correlator $C'(d)$ (solid black line) is fitted by a cosine-modulated Gaussian function (dashed red line) to extract the correlation length $L$. 
	Gray regions are strongly influenced by imaging noise and excluded from fits.  
	Correlators are displayed for $\tho = 50$\,ms (upper) and $\tho = 300$\,ms (lower).  
	{\bf c.} Density-density correlation length $L$ versus $N_c$, for the same timesteps shown in a. 
	Horizontal error bars are the standard deviation over repetitive shots, vertical error bars reflect the correlator fit uncertainty, the red points correspond to the correlators of subfigure b.  
	The dashed line indicates the simple atom-number scaling of the Thomas--Fermi radius of a harmonically trapped BEC, $ \propto N_c^{1/5}.$
	}
	 \label{fig:2} 
\end{figure}

Figure~\ref{fig:2} shows the birth of the supersolid.
Starting from a thermal sample, we apply the ``fast ramp'' (225\,ms) evaporation protocol to the desired final trap depth, too fast for the cloud to follow adiabatically and intermediately resulting in a non-thermalized, non-condensed sample. 
Simply holding the sample at constant trap depth for a time $\tho$, collisions and plain evaporation lead to thermalization and cooling.  
In Figure~\ref{fig:2}a we plot the average axial in-situ density profile (cf.~Fig.~\ref{fig:1}b) versus $\tho$, for about 20~images per time step without any image recentering. 
At early $\tho$ the atoms are primarily thermal, and show up as a broad, low-density background in our images.   
For $\tho \lesssim 150$~ms, inspection of single-shot images reveals an increasing, though substantially fluctuating number of droplets appearing out of the thermal cloud.  
After this time, the droplet number stabilizes to its final value.  
We observe that the droplet formation happens on the same time scale as the equilibration of $N_c$ and $T$ \cite{supmat}, which we expect to be set predominantly by the elastic collision rate. 
Once the droplets have formed, other time scales might be relevant in determining the equilibration rate of their relative positions and phases; the details of this possibility remain an open question~\cite{Ilzhofer:2019}.

To better quantify the growth of the modulated state we consider the density-density correlator $C'(d)$ for the in-situ density profiles over distances $d$~\cite{supmat}.  
We find that $C'(d)$ is well described by a cosine-modulated Gaussian, and define the density correlation length $L$ (Fig.~\ref{fig:2}b) as its fitted width.  
This method provides a way to determine the extent over which density modulation has formed.  
Figure~\ref{fig:2}c shows $L$ for the data set of Fig.~\ref{fig:2}a versus the number of coherent atoms $N_c$, which we extract from TOF absorption images in separate experimental trials with identical parameters.    
Interestingly, despite the strongly modulated structure of the supersolid state, the density correlation length $L$ closely follows a scaling $ \propto N_c^{1/5}$, just as the Thomas--Fermi radius of a harmonically trapped BEC, suggesting a dominant role of interactions over kinetic energy.


\begin{figure}[bht]
    \centering
	\includegraphics[width=\columnwidth]{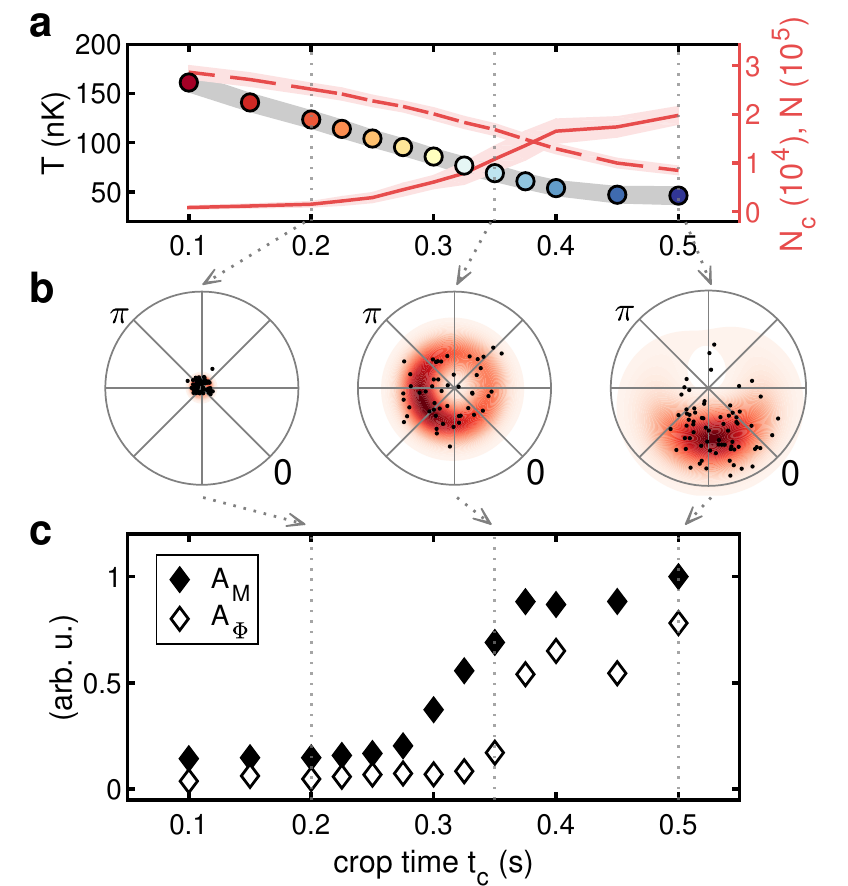}
	\caption{{\bf Development of modulation and coherence while evaporating into the supersolid state.} 
	{\bf a,} Sample temperature $T$ (left ordinate, bullets), total ($N$, right ordinate, dashed red line) and coherent atom number ($N_c$, solid red line) as a function of the ramp crop time $t_c$. The shadings reflect the respective confidence intervals.
	{\bf b,} The phasors $P_i$ (black dots), representing the magnitude and phase coherence of modulation for selected $t_c$ (dotted lines; same radial scale for all polar plots). The red shading reflects mean and variance of the distribution.
	{\bf c,} 
	Evolution of the Fourier amplitude means $\am$ (filled markers) and $\aphi$ (open markers).
	}
	\label{fig:3} 
\end{figure}

While in-situ images provide information about density modulation (diagonal long-range order), they do not carry direct information about phase coherence (off-diagonal long-range order), either within, or between droplets.  
For this, we use TOF imaging and address the question of whether the formation of density modulation precedes global (\ie~interdroplet) phase coherence during evaporative formation of the supersolid, or the other way round.  

For this study, we perform a ``slow" ($500$\,ms) final forced evaporation ramp of constant slope that is nearly adiabatic 
with respect to $N_c$ and $T$ (though not necessarily with respect to excitations of droplet positions and phase), 
and terminate the ramp at selected crop times $t_c$~%
\footnote{ 
While for this protocol the trap parameters do change slightly versus $t_c$, this protocol proved more robust to shot-to-shot fluctuations in atom number than the fast ramp of Fig.~\ref{fig:2}, which is important for this measurement.}.
After $t_c$, we immediately release the atoms and perform TOF imaging.
Figure~\ref{fig:3}a shows the observed evolution of the total ($N$) and (quasi-)condensed ($N_c$) atom number as well as the sample temperature ($T$) versus $t_c$.
We expand on the observed evolution by measuring coherence properties. Following Refs.~\cite{Chomaz:2019,Ilzhofer:2019}, for each measurement $i$ we extract a  rescaled complex phasor $P_i = \rho_i \exp{(-\ii \Phi_i)}$, \ie~the Fourier component corresponding to the modulation wavelength in the TOF interference profile \cite{supmat}.  
For systems with a small number of droplets (but at least two), the magnitude of the phasor $\rho_i$ encodes the modulation strength and also the (local) degree of coherence within each of the individual droplets.  
Meanwhile, the phase $\Phi_i$ depends primarily on the relative phase between the droplets (cf.~\cite{hadzibabic2004interference}).  

We plot the phasors for different evaporation times on the polar plane in Fig.~\ref{fig:3}b, where two effects become apparent.
First, the modulus of the phasors grows during the evaporation, indicating that the degree of modulation increases.  
Second, the distribution of phases $\Phi_i$ is initially uniform, and then narrows down over $t_c$.  
To determine the time sequence of these two effects, we calculate the incoherent and coherent amplitude means,  $\am = \langle|P_i|\rangle_i$, encoding modulation strength and local phase coherence, and $\aphi = |\langle P_i\rangle_i|$, encoding the degree of global phase coherence across the system~\cite{Chomaz:2019,Ilzhofer:2019}.
Plotting $\am$ and $\aphi$ against $t_c$ (Fig.~\ref{fig:3}c), we notice a time lag of around $40$\,ms between the increase of $\am$ and $\aphi$, indicating that during evaporation into a supersolid the translational and the phase symmetry are not broken simultaneously~%
\footnote{We find that observing this time lag does not require a particular fine-tuning of the experimental parameters (starting atom number, $B$-field, trap frequencies, ramp speed) to a higher degree than other experiments with supersolids.
This indicates that this time lag between $\am$ and $\aphi$ is a rather general feature of the evaporation into the supersolid.}.
Rather, density modulation and local phase coherence appear before global phase coherence, consistent with predictions from Monte~Carlo simulations, cf.~\eg~\cite{kora2019patterned}. 
A similar effect is observed in the fast-ramp protocol \cite{supmat}.

This observation suggests the transient formation of a quasi-condensate crystal -- a state with local but not long-range coherence~\cite{Petrov:2001,Dettmer:2001,prokof2004weakly}, whose increased compressibility relative to a thermal gas allows for the formation of density modulation~\footnote{N.~Prokof'ev and B.~Svistunov, University of Massachusetts, Amherst, private discussion.} -- prior to the formation of a supersolid with phase coherence between droplets. 
The lack of global phase coherence could be attributed to a Kibble--Zurek-type mechanism~\cite{damski2010soliton}, in which different regions of the sample condense independently, 
to excitation of modes involving the motion or phase of the droplets during the evaporation process,
or to the thermal population of collective modes (which reduce long-range coherence) at finite temperature.  
As the evaporation process does not allow independent control of temperature and condensation rate without also changing density or trap geometry, we cannot reliably determine the relative importance of these effects (or others) from the experiment.  
Dedicated theoretical studies at finite temperature will thus be needed to elucidate the impact of these types of processes and to understand the exact formation process.


\begin{figure}[ht]
    \centering
	\includegraphics[width=\columnwidth]{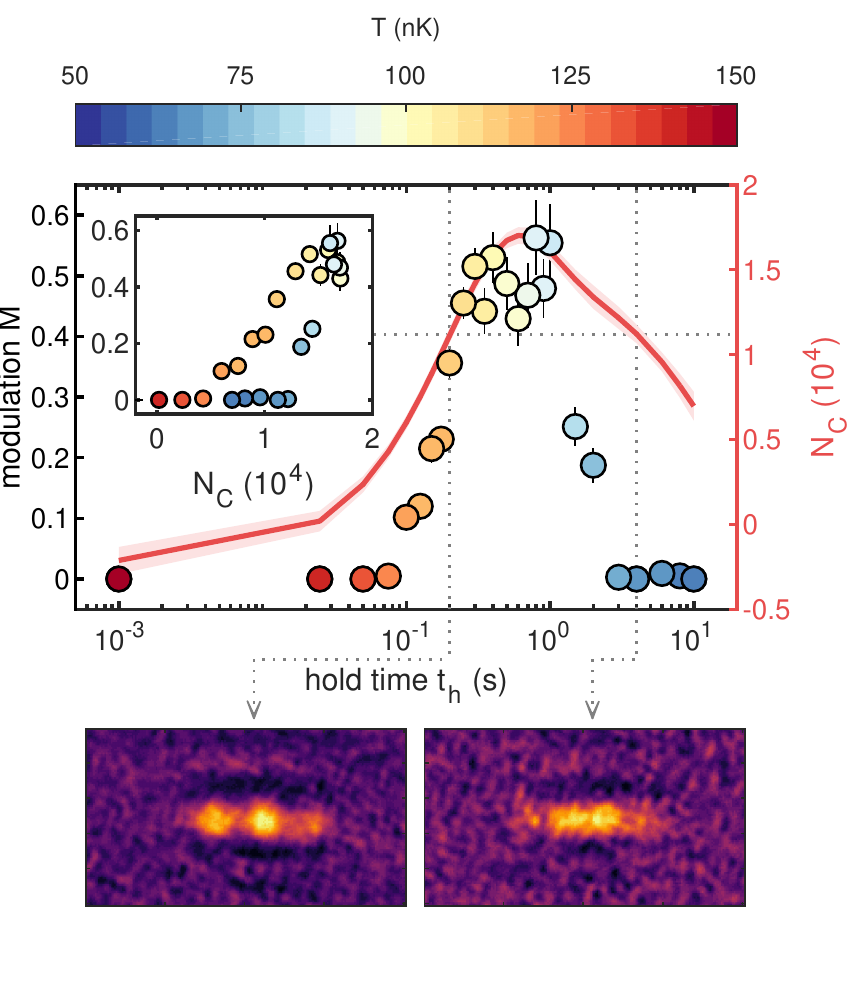}
	\caption {{\bf Lifecycle of a supersolid state.} 
	Density modulation $M$ (from in-situ images) during the evaporation process (left ordinate, bullets;  the vertical error bars reflect the propagated uncertainty returned by the fitting routine). 
	The sample temperature decreases during the hold time $t_h$ and is encoded by the color filling. 
	$N_c$ (from TOF images) is the number of coherent atoms over $t_h$ (right ordinate, red line; the light-red shading reflects the measurement standard deviation).  
	At two times where $N_c \sim 1.1{\times}10^4$ (vertical dashed lines), but at which the atoms have different temperatures, $M$ differs substantially.  
	The corresonding averaged in-situ images below confirm a higher level of modulation at earlier $t_h$.  
	Inset: The observed modulation $M$ plotted versus $N_c$.
	}
	 \label{fig:death} 
\end{figure}

After the birth of the supersolid state, both density modulation and global phase coherence persist for remarkably long times, exceeding one second. 
Figure~\ref{fig:death} shows the evolution of the coherent atom number $N_c$ and temperature $T$ at long hold times under conditions similar to Fig.~\ref{fig:2} 
-- the same fast ramp followed immediately by hold time $t_h$.
Evaporative cooling first increases the coherent atom number until, at long $\tho \geq \SI{1}{\second}$, atom losses become dominant and lead to a continuous decrease of $N_c$, eventually leading to the disappearance of the modulated state. 
However, this death of the supersolid is not a mere time-reversal of the birth. 
$N_c$ decreases, \ie~evolves in the opposite direction, but more slowly and at lower temperature than for the birth. 
Furthermore, phase coherence appears to outlive modulation and to be maintained until the very end~\cite{supmat}.
Thus, a comparison between the birth and death process provides us with important clues to the impact of temperature on the supersolid.

We contrast the birth and death of the supersolid in Fig.~\ref{fig:death} by also plotting the observed in-situ density modulation $M$, which is calculated by Fourier transforming the in-situ density profiles and normalizing the Fourier component corresponding to the modulation wavelength to the zero-frequency Fourier component.  
By comparing $M$ between times that have similar $N_c$ during the birth and the death of the supersolid, respectively, we find that the degree of modulation is higher during the birth of the supersolid than during the death.  
Because the sample is hotter at shorter hold times, this suggests that the observed modulation is increased at higher temperature, perhaps due to thermal population of collective modes, or due to finite-temperature modifications to the dispersion relation~\footnote{Th.~Pohl, University of Aarhus, private communication.}, as predicted in Ref.~\cite{Aybar2019}.
Again, further development of finite-temperature theory will be needed to conclusively determine the importance of such effects.


The role of finite temperature in the formation of modulation, as well as the mechanism by which phase variations across the modulated state arise and then ultimately disappear, represent important future directions for theoretical investigations of dipolar supersolids away from the relatively well understood $T=0$ limit.  
Experimentally, it would be of great interest to study the evaporative formation process in a larger and more uniform system, where distinct domains may be observed to form, and a broader separation of length-scales may be explored in correlation measurements.  Such measurements, along with improved finite-temperature theory, could enable more precise statements as to the nature of the supersolid phase transition away from zero temperature.

\FloatBarrier

\begin{acknowledgments}
We thank Russell Bisset, the Innsbruck Erbium team,  Massimo Boninsegni, Philippe Chomaz, Thomas Pohl, Nikolay Prokof'ev and Boris Svistunov for insightful discussions, and Gianmaria Durastante, Philipp Ilzh\"{o}fer and Arno Trautmann for early contributions. 
This work is financially supported through an ERC Consolidator Grant (RARE, No.\,681432), an NFRI grant (MIRARE, No.\,\"OAW0600) of the Austrian Academy of Science, the QuantERA grant MAQS by the Austrian Science Fund FWF No.\,I4391-N, and the DFG/FWF via FOR~2247/I4317-N36. 
M.S.~acknowledges support by the Austrian Science Fund FWF within the DK-ALM (No.\,W1259-N27).
M.A.N.~has received funding as an ESQ Postdoctoral Fellow from the European Union’s Horizon 2020 research and innovation programme under the Marie Skłodowska‐Curie grant agreement No.\,801110 and the Austrian Federal Ministry of Education, Science and Research (BMBWF). M.J.M.~acknowledges support through an ESQ Discovery Grant, by the Austrian Academy of Sciences. 
L.C.~acknowledges support through the FWF Elise Richter Fellowship No.\,V792. 
We also acknowledge the Innsbruck Laser Core Facility, financed by the Austrian Federal Ministry of Science, Research and Economy.

\end{acknowledgments}

\bibliography{references}




\end{document}